\documentclass{elsart}

\usepackage{graphicx}
\usepackage{epsfig}

\usepackage{amssymb}
\usepackage{color}

\begin{document}

\begin{frontmatter}



\title{Germanium Detector Response to Nuclear Recoils in Searching for Dark Matter}
\author[usd]{D. Barker} and
\author[usd]{D.-M. Mei\corauthref{cor}}
\corauth[cor]{Corresponding author.}
\ead{Dongming.Mei@usd.edu}

\address[usd]{Department of Physics, The University of South Dakota, 
Vermilion, South Dakota 57069}

\begin{abstract}
The discrepancies in claims of experimental evidence in the search for weakly interacting massive particle (WIMP) dark matter necessitate a model for 
ionization efficiency (the quenching factor) at energies below 10 keV. We have carefully studied the physics processes that contribute to the ionization 
efficiency through stopping power. The focus of this work is the construction of a model for the ionization efficiency in germanium by analyzing 
the components of stopping power, specifically that of the nuclear stopping power, at low energies. We find a fraction of the ZBL nuclear stopping power 
can contribute to ionization efficiency. We propose a model that corrects the missing contribution to ionization efficiency from 
the ZBL nuclear stopping power. The proposed model is compared to previous measurements of ionization efficiency in germanium 
as well as that of other theoretical models. Using this new model, 
the thresholds of both CDMS II and CoGeNT are analyzed and compared in terms of the nuclear recoil energy.
  
\end{abstract}
\end{frontmatter}

\maketitle

\section{Introduction}
In the search for dark matter through direct detection of weakly interacting massive particles (WIMPs), it is important to create well defined models for nuclear 
interactions at low energies.  WIMPs are believed to only interact via weak force and gravity; hence the interaction cross section with ordinary matter 
is extremely small. If they are to be detected, it will be from their scatterings off nuclei in the target material. 
 Measuring the recoil of the nuclei (directly or indirectly) will give an estimate of the mass of the incoming WIMPs.

DAMA/NaI~\cite{bern1} and DAMA/LIBRA~\cite{bern2} have observed a model independent, annually modulated signal that has been interpreted as evidence for
 dark matter.  This claim has been met with skepticism within the global dark matter community. Before the end of 2010,
CDMS~\cite{cdms1}, EDELWEISS~\cite{edel1}, 
XENON10~\cite{xenon10}, Xenon100~\cite{xenon100}, and CoGeNT~\cite{CoGeNT2008} published limits that are clearly incompatible with the DAMA signal 
region~\cite{bern1, bern2, dama1} 
by assuming coherent scalar WIMP interactions.  Moreover, the combination of the upper limit contours from the CDMS II~\cite{cdms2}, CRESST I~\cite{cresst1}, 
PICASSO~\cite{pica1}, NAIAD~\cite{naia}, ZEPLIN I~\cite{zep1}, EDELWEISS~\cite{edel2}, SIMPLE~\cite{simp1} and Super-Kamiokande~\cite{superk} experiments 
indicates an inconsistency when interpreting the DAMA signal in terms of a WIMP-nucleon spin-dependent coupling within the standard halo model. A global 
analysis by Fairbairn and Schwetz~\cite{fais} and Savage {\it et al.}~\cite{sava} disfavors the DAMA results as a WIMP-nucleon interaction.  

However, the recent CoGeNT annual modulation signal~\cite{CoGeNT2011}, released in June 2011, is a star of excitement for WIMPs in a low mass range of 7 GeV. 
In response to this excitement, CDMS II~\cite{cdms2011} performed a low threshold analysis 
and found there is no signal in the region of interest that is 
comparable with CoGeNT. On March 5th of 2012, CDMS II released the analysis of annual modulation in the low energy CDMS II data~\cite{cdms2012}
 and found no evidence of annual modulation. Note that both CoGeNT and CDMS II are germanium-type experiments. This discrepancy signifies the importance of 
understanding the detector response to low energy nuclear recoils. 

Since CDMS II has n-$\gamma$ discrimination capability down to the low energy region, and CoGeNT and DAMA do not posses such ability, the possibility that dark 
matter particles interact predominantly with electrons as opposed to nuclei, discussed by many authors including the DAMA 
collaboration~\cite{yas, dpf, mpo, pfa, bern3}, is an interesting approach that is under consideration. This requires some extension of the Standard Model 
but nothing forbids such a possibility. However,
this possibility is now fading because of the most recent analysis of electronic recoil events in CDMS II low energy data~\cite{cdms2012} with no evidence 
of annual modulation. In addition, the result from CRESST-II~\cite{cresst2011}, which claimed the excess of events in their energy range that 
indicates a WIMP mass of 13 GeV, differs from CoGeNT's mass of WIMPs. 
 CRESST-II does possess a good n-$\gamma$ discrimination in their energy region of interest. This is  another indication of the importance 
of understanding the detector response to low energy nuclear recoils. Note that Xenon100 shows no evidence of WIMP mass 
in the region of 7 GeV to 13 GeV~\cite{xenon100}. There is also an  
alternative approach to the explanation of experimental results assuming an O-helium universe~\cite{maxim}.

Lack of a coherent picture to explain the results from DAMA, CoGeNT, and CRESST-II requires more attention be paid to the interpretation of the 
experimental results, such as the ionization efficiency, which is critical to the detection threshold of energy, hence the region of interest and the WIMP 
masses.  Since CoGeNT and CDMS II are both germanium detectors and the results conflict with each other, it is necessary to study the ionization efficiency of 
germanium detectors.  Several theoretical models exist for ionization efficiency, however the best model for interpreting the experimental data has yet
 to be established. Many experiments adopt a model proposed by Lindhard {\it et al.} in the 1960s~\cite{lind} with different values of a constant, $k$, 
in their data analysis. For
example, within the limits of $k=0.1$ to $k=0.2$, $k=0.15$~\cite{kwj, cch, tsh, lba}, $k=0.157$~\cite{yme, bse}, $k=0.159$~\cite{ars} and 
$k=0.2$~\cite{kwj, psb} were used in the interpretation of the experimental data. The difference in the ionization
efficiency given by different values of $k$ can be as large as 30\%. Without knowing which value of $k$ is correct in terms of physics in the low energy region, 
it is difficult to justify the threshold of nuclear recoil energy and the mass of WIMPs. Therefore, a model that demonstrates physics processes in the low 
energy region is needed. This model will also justify which value of $k$ should be used in Lindhard's model.
 The model presented in this paper has been developed by re-examining the components of stopping power, especially that of the nuclear 
stopping power, at low energies.

\section{Developing the Model for Ionization Efficiency}
Previous work by Mei {\it et. al.}~\cite{dmm} developed models for the quenching factor in noble liquids.  This work modifies the previous 
work to look
 at the ionization efficiency in germanium, taking another look at the stopping power of the germanium ions produced in the WIMP-nucleus 
interaction in germanium.
\subsection{Electronic Stopping Power} 
\begin{figure}[htb!!!]
\includegraphics[angle=0,width=12.cm] {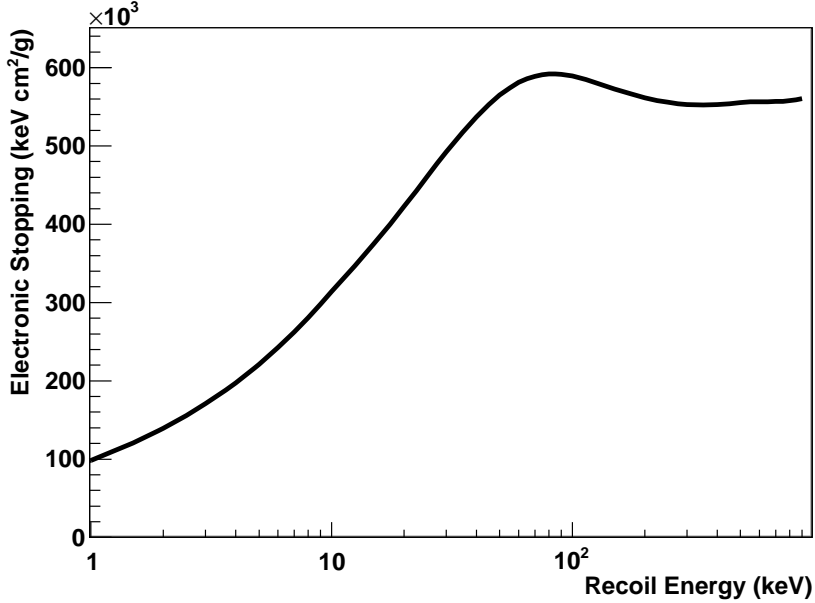}
\caption{\small{
Shown is the calculated electronic stopping power in germanium for the low energy region of nuclear recoils.}}
\label{fig:electronicstoppingpower}
\end{figure}

The electronic stopping power is well understood as the orbital electrons of the surrounding atoms that are ionized by inelastic collisions with the recoiling ion 
along its path. This portion of energy loss by a recoiling nucleus contributes to the ionization efficiency at a level of 100\%. 
 To calculate the electronic 
stopping power, the “heavy ion scaling rule” was applied to calculate the stopping power of germanium ions in germanium from the stopping power of 
protons in germanium at the same velocity:
\begin{equation}
S_{Ge} = (\xi Z_{Ge})^2S_p.
\end{equation}
The effective charge fraction, $\xi$, was approximated by Brand and Kitagawa~\cite{brk} and modified by ZBL~\cite{zbl} as
\begin{equation}
\xi = q + 0.5\cdot(1-q)\frac{v_{0}}{v_{f}}\cdot ln\left[1+\left(\frac{4\lambda \cdot 2.02}{1.919}\right)^2\right],
\end{equation}
where $v_F$ is the Fermi velocity, $v_0$ the Bohr velocity, $\lambda$ is given by
\begin{equation}
\lambda = \frac{0.48(1-q)^{\frac{2}{3}}}{Z_{Ge}^{\frac{1}{3}}}\cdot a_{0}\left[1-\frac{(1-q)}{7}\right],
\end{equation}
with $q$ given by
\begin{equation}
q = 1-exp\left[0.0373292y_{r}^{0.3}+0.0571791y_{r}^{0.6}-1.09793y_{r}+0.072649y_{r}^2\right],
\end{equation}
and $a_{0}$ is the Bohr radius. $y_{r}$ is a dimensionless quantity related to the relative velocity by
\begin{equation}
y_r = \frac{v_{r}}{v_{0}\cdot Z_{Ge}^{\frac{2}{3}}}.
\end{equation}
The relative velocity is given in~\cite{amw}, and is related to $v_{F}$ and the ion velocity, $v_{1}$, by
\begin{equation}
v_r = \left\{ \begin{array}{ll} 
\frac{3}{4}v_F\left(1+\frac{2}{3}\frac{v_1^2}{v_F^2}-\frac{1}{15},
\frac{v_1^4}{v_F^4}\right), & \textrm{if $v_1 \leq v_F$}, \\
v_1\left(1+\frac{1}{5}\frac{v_F^2}{v_1^2}\right), & \textrm{if $v_1 > v_F$} . 
\end{array}\right.
\end{equation}
Figure~\ref{fig:electronicstoppingpower} shows the calculated electronic stopping power in germanium.

\subsection{Nuclear Stopping Power}
Nuclear stopping power is less understood at the low energies induced by the recoiling ion in a WIMP collision.
Nuclear stopping represents elastic collisions between the recoiling ion and the surrounding atoms in the target. If one knows the form of the 
repulsive potential between them, it is possible to calculate the nuclear stopping power.  
The general calculation of nuclear stopping power is performed with the following equations from ZBL~\cite{zbl}:
\begin{equation}
\label{nstop}
S_{n}(E) = \frac{8.462\cdot 10^{-15}Z_{1}Z_{2}M_{1}S_{n}(\epsilon)}{(M_{1}+M_{2})(Z_{1}^{0.23}+Z_{2}^{0.23})},
\end{equation}
where M$_{1}$ and M$_{2}$ are the projectile and target masses (amu), and Z$_{1}$ and Z$_{2}$ are the projectile and target atomic
numbers, respectively. $\epsilon$ is the reduced energy given by
\begin{equation}
\epsilon = \frac{32.53M_{2}E}{Z_{1}Z_{2}(M_{1}+M_{2})(Z_{1}^{0.23}+Z_{2}^{0.23})},
\end{equation}
and the reduced nuclear stopping power
\begin{equation}
\left\{ \begin{array}{ll}
S_{n}(\epsilon)=\frac{ln(1+1.1383\epsilon)}{2(\epsilon+0.01321\epsilon^{0.21226}+0.19593\epsilon^{0.5})}, & \textrm for \epsilon\leq30, \\
S_{n}(\epsilon)=\frac{ln(\epsilon)}{2\epsilon}, & \textrm for \epsilon>30.
  \end{array}\right.
\end{equation}
Note $E$ is kinetic energy of projectile in keV.

\begin{figure}[htb!!!]
\includegraphics[angle=0,width=12.cm] {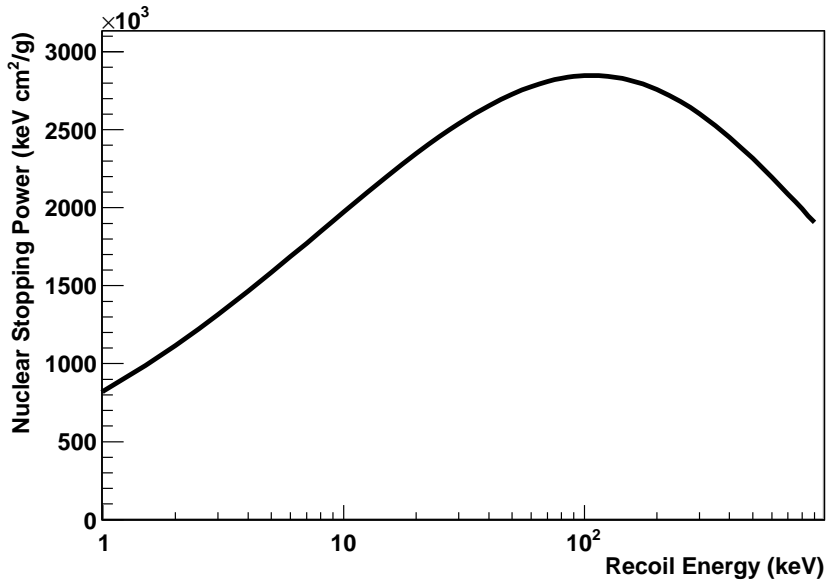}
\caption{\small{
Shown is the calculated nuclear stopping power in germanium for the low energy region of nuclear recoils. }}
\label{fig:nuclearstoppingpower}
\end{figure}

We show the calculated nuclear stopping power in Figure~\ref{fig:nuclearstoppingpower} using the above formulas.
 
In general, this part of energy loss
does not contribute to ionization efficiency. However, the standard deviation of the fit of the universal ZBL repulsive potential to 
the theoretically calculated potentials is 18\% above 2 eV~\cite{zbl}. This indicates that nuclear stopping power calculated with the ZBL repulsive 
potential possesses large uncertainty. Comparing with a more accurate repulsive potential generated from self-consistent total energy calculations, 
the calculation of nuclear stopping power using ZBL repulsive potential cannot be treated totally as the nonionizing nuclear stopping power. Considering 
density-functional theory, the local-density approximation (LDA) for electronic exchange and correlating both the 
electronic part and the inter-nuclear Coulomb repulsion as a function of the distance between atoms in a dimer bond, it became obvious to us that a 
correction needed to be derived to account for the fraction that contributes to ionization efficiency~\cite{kai, jak}.

\subsection{Correction Factor}
\begin{figure}[htb!!!]
\includegraphics[angle=0,width=12.cm] {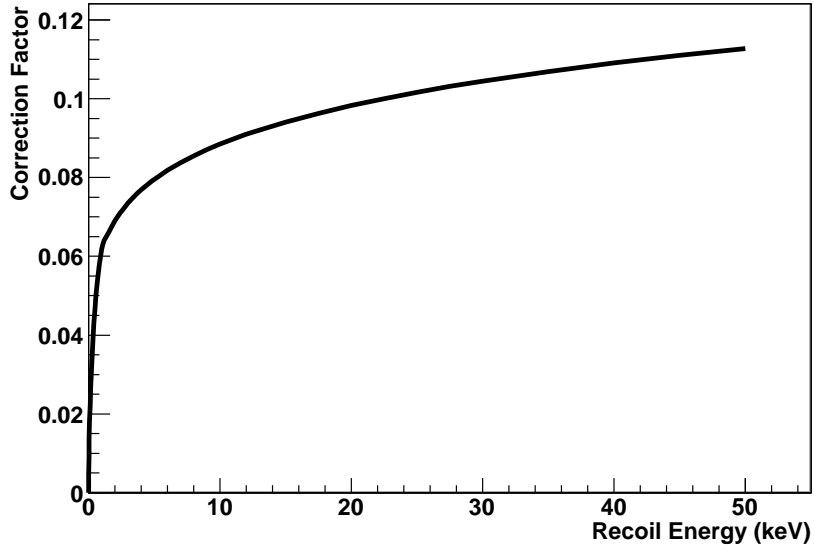}
\caption{\small{
The fraction of the ZBL nuclear stopping power contributing to the ionization efficiency as a function of nuclear recoil energy. }}
\label{fig:fraction}
\end{figure}

When atoms receive significant recoil energies, they will be removed from their lattice positions, and produce a cascade of further collisions in the lattice. 
In the low energy range, the most prominent component of nuclear stopping power is the non-ionizing nuclear stopping power that describes the rate of energy 
loss due to atomic displacements as a recoil nucleus traverses a material~\cite{mjb, iju, jlm}. In the calculated nuclear stopping power using
the ZBL repulsive potential, there is a small 
fraction of the nuclear stopping power that contributes to the calculation of ionization efficiency. It is the goal of this model to incorporate 
this fraction to develop a more accurate model of the ionization efficiency.
The calculation of the fraction of the ZBL nuclear stopping power that contributes to the ionization efficiency is performed by the calculation of
the non-ionizing energy-loss (NIEL) using the Wentzel-Moli\`{e}re differential cross section discussed in Section 2 [Eq. (15)] of reference~\cite{mjb}.
The NIEL can be calculated with
\begin{equation}
\label{niel}
-\left(\frac{dE}{dx}\right)_{nucl}^{NIEL}=n_{A}\int_{E_{r}(d)}^{E_{r}(max)}E_{r}L(E_{r})\frac{d\sigma^{WM}(E_{r})}{dE_{r}}dE_{r},
\end{equation}
where $E$ is the kinetic energy of the incoming particle (in keV), $E_{r}$ is recoil energy (the kinetic energy transferred to the target atom) of the atom 
(in keV), and 
L($E_{r}$) is a fraction of the recoil energy, $E_{r}$, which undergoes displacement processes. The expression of L($E_{r}$) is often called Lindhard 
partition function,
 which can be found 
in Ref.\cite{ijun, srm} and in Equations
(4.94, 4.96) of Section 4.2.1.1 in Ref.\cite{cle}. $E_{r}(de)$ = $E_{r}L(E_{r})$ is the damage energy, i.e., the energy deposited by a recoil nucleus
with kinetic energy $E_{r}$ undergoing displacement damage inside the medium. The integration lower limit of $E_{r}(d)$ = 23 eV in germanium~\cite{hkk} is the
 energy necessary
to displace the atom from its lattice position. The upper limit of $E_{r}(max)$ is the maximum energy that can be transferred during a single collision process. 
Taking the difference between Eq.~\ref{nstop} and Eq.~\ref{niel} divided by the nuclear stopping power (Eq.~\ref{nstop}) gives the fraction of the nuclear 
stopping power that contributes to the ionization efficiency. This fraction of the ZBL nuclear stopping power was found to be
\begin{equation}
C_{f} = 6.2\times10^{-2}E_{r}^{0.15},
\end{equation}
where $E_{r}$ is the recoil energy of the target atom received from the incoming particle.
Figure~\ref{fig:fraction} shows the fraction of the ZBL nuclear stopping power that contributes to the ionization efficiency.

\subsection{Ionization Efficiency}
After calculating the fraction of the ZBL nuclear stopping power that contributes to the ionization efficiency,  it is applied to the ionization efficiency 
in the following way:
\begin{equation}
\varepsilon_{c} = \frac{\eta(E_{r})+C_{f}\nu(E_{r})}{\eta(E_{r})+\nu(E_{r})},
\end{equation}
where $\eta(E_{r})$ is the electronic stopping power, $\nu(E_{r})$ is the ZBL nuclear stopping power and $C_{f}$ is the correction fraction.
  
\begin{figure}[htb!!!]
\includegraphics[angle=0,width=12.cm] {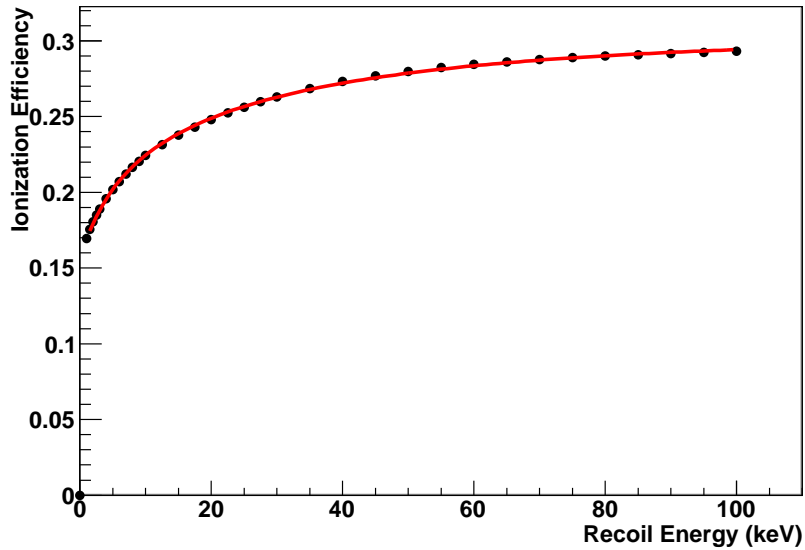}
\caption{\small{
The calculated  ionization efficiency as a function of nuclear recoil energy. }}
\label{fig:efficiency}
\end{figure}

With this fraction of the ZBL nuclear stopping power included in the ionization efficiency, the theoretical model will be a more accurate picture of
the experimental signal. Figure~\ref{fig:efficiency} shows the calculated ionization efficiency as a function of nuclear recoil energy.

The fitted function can be expressed as:
\begin{equation}
\varepsilon_{c} = \frac{0.14476\cdot E_{r}^{0.697747}}{-1.8728+exp[E_{r}^{0.211349}]}.
\end{equation}
This equation is valid for recoil energy, $E_{r}$, greater than 1 keV and less than 100 keV. The corresponding visible energy, $E_{v}$, can be obtained 
by multiplying the ionization 
efficiency by the recoil energy.

It was important to ascertain whether nuclear recoils off the naturally occurring isotopes of germanium in the detectors could produce large differences in the 
ionization efficiency. Calculations were made for each natural isotope of germanium (70, 72, 73, 74, 76) and it was determined that the spread in ionization 
efficiency was less than 2\%. This will contribute a difference of no more than a $\pm$0.04 keV change to the visible energy for a recoil energy of 2 keV.
Therefore, the average atomic mass was used in all calculations.

\section{Comparison to the Experimental Data}
\begin{figure}[htb!!!]
\includegraphics[angle=0,width=12.cm] {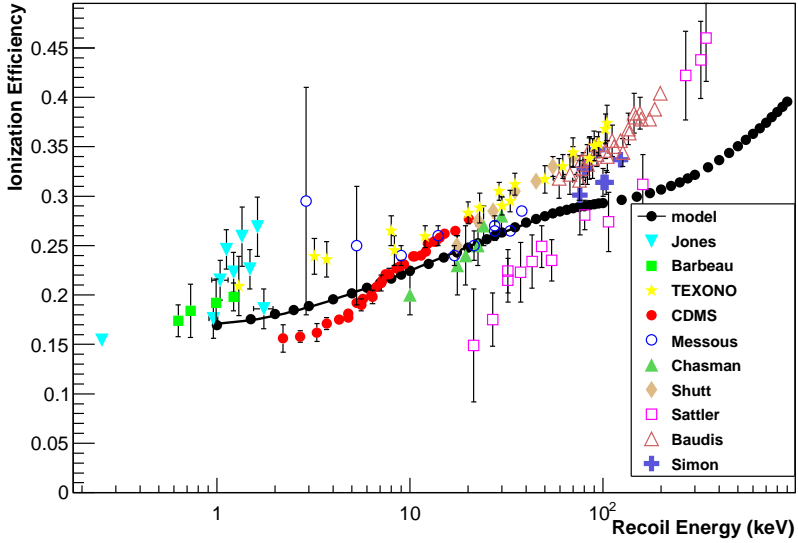}
\caption{\small{
A comparison between the calculated and the measured  ionization efficiencies. Note that the measured data points are from Jones {\it et al.}~\cite{kwj, kwj1}, 
Barbeau {\it et. al.}~\cite{psb}, TEXONO~\cite{TEXONO}, CDMS~\cite{cdms2011}, Messous {\it et al.}~\cite{yme}, Chasman {\it et al.}~\cite{cch}, 
Shutt {\it et al.}~\cite{tsh}, Sattler {\it et al.}~\cite{ars}, Baudis {\it et al.}~\cite{lba}, and Simon {\it et al.}~\cite{esi}.}}
\label{fig:com1}
\end{figure}

Previous works have measured the relation between ionization efficiency or visible energy and the recoil energy in 
germanium~\cite{kwj, cch, tsh, lba, yme, ars, psb, kwj1, TEXONO, esi}.  The calculated ionization efficiency in this work is compared to the measured 
data points from various measurements in 
Figure~\ref{fig:com1}. 
As can be seen in Figure~\ref{fig:com1}, the measured data points from different experiments do not entirely agree with each other. This indicates 
the systematic errors involved in different 
measurements. 

Depending on the experimental techniques, the systematic errors can be dominated by different sources.  For instance, thermal neutrons can be used to measure 
nuclear recoils down to sub-keV region. In this measurement, unless the out-going gamma rays are fully absorbed and measured by another detector in coincidence, 
Compton scattering of the out-going gamma-ray inside the germanium detector can contaminate the visible energy. In the case of measuring nuclear recoils with 
elastic scattering, the multiple scatters must be excluded by a Monte Carlo simulation. In addition, the scattering angle and the time of flight of the out-going 
neutrons have to be measured very accurately.  However, there is a limitation on how fast the electronics can possibly respond to the signal.  If one uses inelastic scattering 
to measure nuclear recoils, the de-excitation of gamma rays through Compton scattering in the detector can contaminate the signal.  Therefore, all of the 
measurements must be implemented together with a reliable Monte Carlo simulation, which is well calibrated by gamma rays in the same energy range, to reduce the 
systematic error and deliver reliable ionization efficiency.  This is to say that a best fit to some of data points in the energy region of interest will not 
necessarily represent the true ionization efficiency.  The physics that involves the ionization efficiency must be understood.

Because of a large spreading of experimental data points, the predicted ionization efficiency from the model agrees with some data points but disagrees with other data points, 
with the majority of measurements 
above the model.

The CDMS II data was collected comparing the ionization yield to the recoil energy~\cite{cdms2011}. The possible discrepancy could be due to 
the lack of the calibration of the 
phonons in the low energy region. 

\section{Comparison to the Previous Models}
The most accepted model for calculating the ionization efficiency, $\varepsilon$, of any element is developed by Lindhard {\it et. al.}~\cite{lind}
\begin{equation}
\varepsilon = \frac{k\cdot g(\epsilon)}{1+k\cdot g(\epsilon)},
\end{equation}
where $\epsilon$ = 11.5$E_{r} Z^{-7/3}$ for a given atomic number, $Z$, $k$ = 0.133$Z^{2/3}A^{-1/2}$, and $g(\epsilon)$ is fitted by 
$g(\epsilon) = 3\epsilon^{0.15} + 0.7\epsilon^{0.6} + \epsilon$.  

\begin{figure}[htb!!!]
\includegraphics[angle=0,width=12.cm] {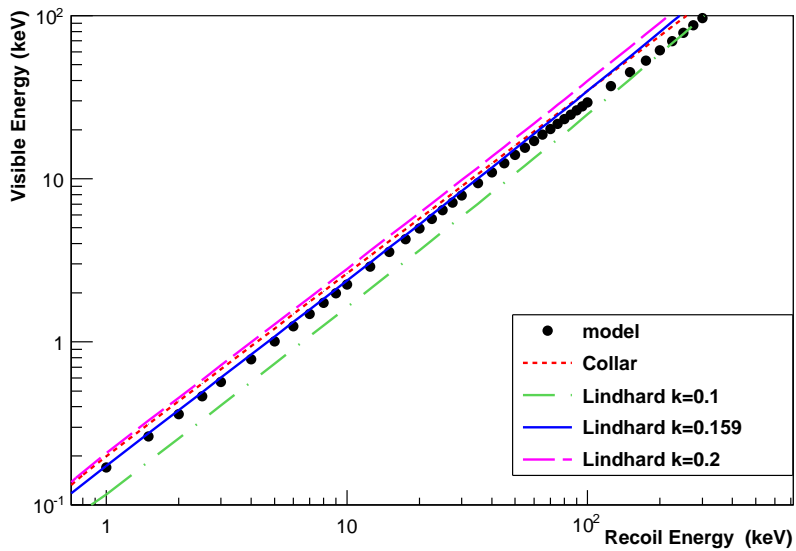}
\caption{\small{
Visible energy versus nuclear recoil energy from different models. }}
\label{fig:model}
\end{figure}

Often two different values of the scaling constant have been used in conjunction with the 
experimental data, one that agrees better at the higher recoil energies and one at the lower.  These are plotted against the model for $k=0.1$ and $k=0.2$. 
The calculated 
value of $k$ for germanium is $k=0.159$.  This is plotted with the traditional $k$ values in Figure~\ref{fig:model}. It is interesting to note that 
at low energies 
$k=0.159$ agrees well with the model proposed by this work.  Thus, if Lindhard's model is to be used, the proper value of $k$ for low energies 
is, $k=0.159$, for germanium.  
Also note that at higher energies the proposed model tends to agree with the Lindhard model with $k=0.1$.

A best fit to the data between 1 keV and 10 keV has been used by J.I. Collar {\it et al.}~\cite{psb, jic} and others~\cite{cdms2012, tsj} in order to interpret the 
CoGeNT and CDMS II results:
\begin{equation}
\label{Collar}
E_{v} = 0.199E_{r}^{1.12}.
\end{equation}
This best fit appears to follow the Lindhard model $k=0.2$ at low energies and deviate to 
become parallel with Lindhard $k=0.1$ 
as the energy increases.

Figure~\ref{fig:model} shows a comparison between the proposed model, Lindhard's model, and Collar's best fit.

\section{Validation of Models with Experimental Data}
\begin{figure}[htb!!!]
\includegraphics[angle=0,width=12.cm] {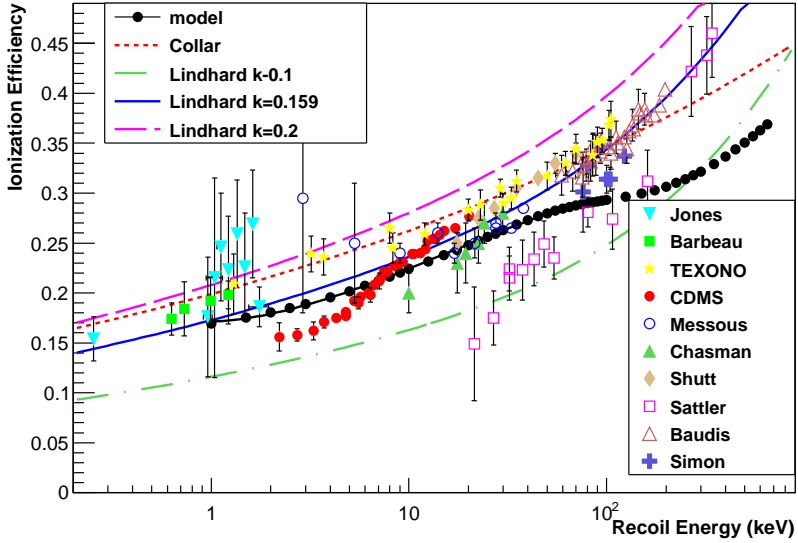}
\caption{\small{
Ionization efficiency as a function of energy of nuclear recoils. Shown is a comparison between the measured data points and the popular models including 
the proposed model in this work. }}
\label{fig:vali}
\end{figure}

Both CDMS II and CoGeNT are low threshold detectors accessible to low energy recoils induced by low mass WIMPs. 
Therefore, the ionization efficiency in the low energy region is critical to both experiments. To validate a reliable model for ionization
efficiency in the low energy region, we plot the experimental data points together with several theoretical models including the proposed 
model in this work in Figure~\ref{fig:vali}.  
It is quite clear in Figure~\ref{fig:vali} that Lindhard's model with $k = 0.1$ or $k = 0.2$ does not agree with the measured data points in particular for the low 
energy region. This is because $k = 0.159$ for germanium according to Lindhard's theory~\cite{lind}. There is no theoretical ground to choose $k = 0.1$ or
$k = 0.2$ in Lindhard's model. If $k = 0.159$ is chosen for Lindhard's model, there is a fair agreement between Lindhard's model and the proposed 
model in this paper. The difference is on the average of 10\% across the energy range from 1 to 100 keV. This can be understood according to Lindhard's theory
in which Lindhard pointed out that his model possesses uncertainty in the low energy region~\cite{lind14}. It is thus important to notice that $k = 0.159$ 
must be chosen 
if one uses Lindhard's model for germanium detector. The 10\% error could shift the analysis threshold by $\pm$0.2 keV at the recoil 
energy of 2 keV. This is a large enough error to generate argument and result in discrepancy in the analysis. A numerical comparison between different
values of constants for Lindhard's model, the best fit of Collar {\it et al.}, and the proposed model is shown in Table~\ref{tab:comp}. As can be seen in 
Figure~\ref{fig:vali} and Table~\ref{tab:comp}, the proposed model is reliable from 1 keV to 100 keV.  

\begin{table}[htb!!!]
\caption{ Comparison between visible energies at a given recoil energy for the models and the Collar's best fit (all energies in keV).}
\label{tab:comp}
\begin{tabular}{|l|l|l|l|l|l|}
\hline \hline
 & & & \multicolumn{3}{|c|}{Lindhard}\\
\hline
Recoil Energy & Proposed Model & Collar & k = 0.1 & k = 0.159 & k =0.2\\
\hline
1 & 0.169 & 0.199 & 0.116 & 0.174 & 0.208\\
2 & 0.361 & 0.433 & 0.256 & 0.381 & 0.455\\
5 & 1.01 & 1.21 & 0.733 & 1.08 & 1.28\\
10 & 2.24 & 2.62 & 1.63 & 2.37 & 2.80\\
20 & 4.96 & 5.70 & 3.65 & 5.26 & 6.17\\
50 & 14.0 & 15.9 & 10.7 & 15.2 & 17.7\\
\hline \hline
\end{tabular}
\end{table}

\section{Application of the Proposed Model to the Thresholds of CDMS II and CoGeNT}
CoGeNT claimed the evidence of annual modulation indicating a mass of $\sim$7 GeV WIMPs~\cite{CoGeNT2011}. CDMS II failed to confirm 
CoGeNT's claim~\cite{cdms2012}. Both experiments use germanium detectors located in the Soudan Mine. CoGeNT cannot discriminate electronic recoils 
from nuclear recoils while CDMS II possesses a good capability of discriminating n-$\gamma$ events. The discrepancy between two experiments lies in whether
CoGeNT sees nuclear recoil events that are below the detection threshold of CDMS II.

CDMS II used a 2 keV nuclear recoil threshold in the analysis of the 
low energy region~\cite{cdms2011} in which the ionization efficiency is 15.8\%~\cite{cdms2011}. 
We calculate a difference in the energy scale by $\frac{0.1805 - 0.158}{0.1805}$ = 12.5\%, where 18.05\% is the ionization efficiency at 2 keV nuclear 
recoil energy from the proposed model.  
To interpret the CDMS II threshold with the proposed model, we multiply it by the ratio of the ionization efficiencies: 
 2 keV$\times \frac{0.158}{0.185}$ = 1.75 keV. This is the CDMS II threshold in the energy scale of the proposed model
assuming that the CDMS II energy scale determined by the phonon signal is correct.

\begin{figure}[htb!!!]
\includegraphics[angle=0,width=12.cm] {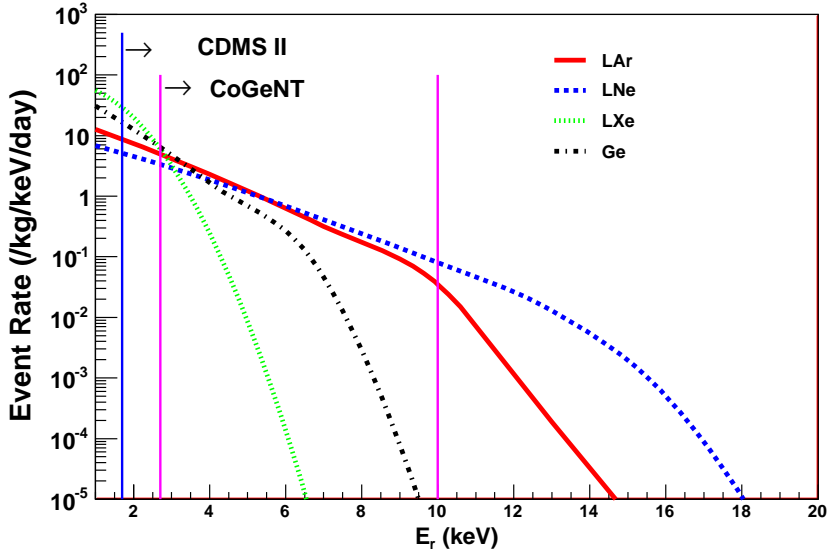}
\caption{\small{
The threshold energy of CoGeNT versus that of CDMS II for WIMPs with mass of 7 GeV and an interaction cross section of 2.5 $\times 10^{-41}$ cm$^{2}$. Included in this plot is the event rate verses recoil energy of other common (or considered)
 targets for WIMP dark matter detection.}}
\label{fig:CoGeNTCDMS}
\end{figure}

Using the proposed model, the analysis threshold of CoGeNT, 0.5 keV electronic equivalent energy, is calculated to be 2.7 keV nuclear recoil energy. 
The lowest threshold
of CoGeNT, 0.4 keV electronic equivalent energy, corresponds to 2.23 keV of nuclear recoil energy. Thus, the CDMS II threshold of 1.75 keV is 
lower than both the 2.7 keV analysis threshold of CoGeNT and the 2.23 keV lower limit of the CoGeNT detector.  Figure~\ref{fig:CoGeNTCDMS} shows a 
threshold energy comparison between
CoGeNT and CDMS II. As can be seen, the CoGeNT is fully contained by CDMS II using our model in this paper.

Using Collar's best fit equation, Eq.~\ref{Collar}, the lowest electronic equivalent energy of 0.4 keV corresponds to 1.87 keV of nuclear recoil. 
The difference between the Collar's best fit equation and the model proposed in this work is about 18\% in calculating nuclear recoil energy for a visible 
energy of 0.4 keV.

\section{Conclusion}
We have developed a new model for ionization efficiency by analyzing the components of stopping power and the fundamental physics 
that must be considered to accurately 
understand low energy nuclear stopping power.  This model is first calculated for germanium as it is the simplest of detectors. The result is compared to both 
experimental data and previous theoretical models.  When compared with the experimental data points, this model passes in the middle of the data points. 
 It shows some agreement with a few sets of data but discrepancies between others. Many sets of experimental data did not span the entire range needed, 
especially at low energies data points are scarce, and so verification will be necessary as a 
next step.  When compared to the theoretical models, there is a fair agreement between Lindhard model with  $k=0.159$ and our model at low energies. 
This value of 
$k$ is the calculated value for germanium ($Z=32$, $A=72.64$) in the Lindhard theory.  

The analysis threshold of the CoGeNT detector is 2.7 keV nuclear recoil energy using the proposed model in this paper,  which is higher than that of the  
1.75 keV threshold of the CDMS II detector in the same model.  
Thus CoGeNT’s region of interest 
in completely contained within that of the CDMS II detector. 
Based on the analytic model developed here and the published results of the two collaborations~\cite{CoGeNT2011, cdms2011}, 
the excess events above the background seen by CoGeNT seem to not be nuclear recoils.

Work will continue on the proposed model to verify its accuracy in the range of 1 keV to 10 keV.  Understanding the ionization efficiency in germanium is 
the first step to building a set of models for ionization efficiency in other materials employed by dark matter detectors. This will provide reliable analytical
 models for the interpretation of experimentally 
observed signals.

\section*{Acknowledgments}
The authors wish to thanks to the research group at University of
South Dakota for the invaluable support that made this work successful. We also wish to thank Matthew Szydagis, Christina Keller, 
and Angela A. Chiller for a careful reading of this manuscript. Thanks to J.I. Collar and Dave Moore for their useful discussion after reading the first version.
This work was supported by the NSF grant 0758120 as well as the DOE grant DE-FG02-10ER46709 and the state of South Dakota.

%
%

\end{document}